\documentclass[12pt]{article}

\usepackage[body={17.5cm, 22.5cm},right=2cm]{geometry}

\usepackage{color}
\usepackage{graphicx}
\usepackage{hyperref}
\usepackage{epsf}
\usepackage{graphicx,epsfig}
\usepackage[skip=2pt,font=small]{caption}

\usepackage{float}

\usepackage{amsmath}
\usepackage{amssymb}
\usepackage{epsfig}
\usepackage{cite}
\usepackage{color,colordvi}
\newcommand{\be}{\begin{eqnarray}}
\newcommand{\ee}{\end{eqnarray}}
\newcommand{\bi}{\begin{itemize}}
\newcommand{\ei}{\end{itemize}}

\newcounter{hran}

\def\MSbar{\relax\ifmmode\overline{\rm MS}\else{$\overline{\rm MS}${ }}\fi}

\def\n3{\nu_3}

\def\n1{\nu_1}

\numberwithin{equation}{section}
\begin{document}

\begin{flushright}
LMU-ASC 26/15, MPP-2015-95, CERN-PH-TH-2015-099,
May 2015
\vspace{-.0cm}
\end{flushright}

\thispagestyle{empty}

\vspace*{1cm}
\begin{center}

\def\thefootnote{\fnsymbol{footnote}}

{\LARGE \bf 
Aspects of Quadratic Gravity
}
\\[1.5cm]
{\large  \bf Luis Alvarez-Gaume$^{a}$, Alex Kehagias$^{b,c}$, 
Costas Kounnas$^{d}$, \\[0.3cm]
Dieter L\"ust$^{e,f,a}$  and Antonio Riotto$^{c}$}
\\[0.5cm]

\vspace{.3cm}
{\normalsize { \it $^{a}$
CERN, PH-TH Division\\
1211 Geneva 23,
Switzerland}}
\\
\vspace{.3cm}
{\normalsize {\it  $^{b}$ Physics Division, National Technical University of Athens, \\15780 Zografou Campus, Athens, Greece}}\\

\vspace{.3cm}
{\normalsize { \it $^{c}$ Department of Theoretical Physics and Center for Astroparticle Physics (CAP)\\ 24 quai E. Ansermet, CH-1211 Geneva 4, Switzerland}}\\

 \vspace{.3cm}
{\normalsize { \it $^{d}$
Laboratoire de Physique Th\'eorique, Ecole Normale Sup\'erieure,\\
24 rue Lhomond, 75231 Paris cedex 05, France}}\\

\vspace{.3cm}
{\normalsize { \it $^{e}$
Max-Planck-Institut f\"ur Physik (Werner-Heisenberg-Institut),\\
F\"ohringer Ring 6, 80805 M\"unchen, Germany}}\\

\vspace{.3cm}
{\normalsize { \it $^{f}$
Arnold Sommerfeld Center for Theoretical Physics,\\
LMU, 
Theresienstr. 37, 80333 M\"unchen, Germany}}

\vspace{.3cm}


\end{center}

\vspace{.5cm}

\hrule \vspace{0.3cm}
{\small  \noindent \textbf{Abstract} \\[0.3cm]
\noindent 
We discuss quadratic  gravity where terms quadratic in the curvature tensor are included in the  action.  
After reviewing the corresponding field equations, we analyze in detail  the physical propagating modes in some specific backgrounds.  
First we confirm that  the pure $R^2$ theory is indeed ghost free. Then we point out that 
 for flat backgrounds the pure $R^2$ theory propagates only a  scalar massless mode  and no spin-two tensor mode. However, the latter  emerges either by expanding
 the theory around curved backgrounds like de Sitter or anti-de Sitter, or by changing the long-distance dynamics by introducing the standard Einstein term. In both cases, the theory
 is modified in the infrared and a propagating graviton is recovered. Hence we recognize a subtle interplay between the UV and IR properties of
 higher order gravity.
 We 
also calculate  the corresponding Newton's law for general quadratic curvature theories. Finally, we discuss how quadratic actions may be obtained from a fundamental theory like string- or M-theory. We demonstrate that string theory on  non-compact $CY_3$ manifolds, like a  line bundle over $\mathbb{CP}^2$,
may indeed lead to gravity dynamics determined by a higher curvature action. 

\vspace{0.5cm}  \hrule
\vskip 1cm

\def\thefootnote{\arabic{footnote}}
\setcounter{footnote}{0}


\baselineskip= 15pt

\tableofcontents

\vskip2cm

\section{Introduction}

There is recently a renewed interest in higher curvature theories. These are theories of the general form $R+{\cal  R}^n$, where ${R}$ is the standard Einstein gravity and ${\cal R}^n$ denotes collectively $n^{\rm th}$ power of the Riemann, Ricci, Weyl tensors or the curvature scalar \cite{Stelle,Boulware,David, Horowitz,Deser,tHooft,Maldacena}. Of course, in string theory we are familiar with such a structure as the string effective action contains an infinite,  well organized and ghost-free series of higher curvature corrections to the leading  Einstein gravity. In particular, as has been noted in \cite{Stelle}, the inclusion of quadratic curvature terms in the  action makes the theory  renormalizable.  Contrary to  string theory, in the truncated, effective field theory  there is a price to be  paid namely  the appearance of a massive ghost state. This ghost state originates from the square of the  Riemann,  the Ricci or the Weyl tensor. 
Note that, as we will discuss, 
if only the  scalar curvature and its square are included in the gravity action, then, contrary to some people's belief, these theories are indeed physical and do not contain ghost like modes.  In fact, the  $R+R^2$ theory, known as the Starobinsky model \cite{Starobinsky}, propagates besides the usual massless graviton, an additional massive spin-0 state,  known as the ``scalaron field"  or  the so called ``no-scale field".  
After a field redefinition \cite{Whitt}, one obtains
 a scalar field  minimally coupled to standard Einstein gravity  with a  potential
making the $R+R^2$ theory particularly appealing for cosmological inflation \cite{Guth}.
The $R+R^2$ theory can also be embedded in supergravity \cite{Cecotti}, and 
there is a large amount of recent work on the inflationary predictions of the supersymmetric  $R+{ R}^2$ theory
 \cite{Ellis:2013xoa,Kounnas:2014gda}. In addition, quadratic gravity theories have also been discussed in particle physics on the basis of their properties under scale transformations.
\cite{Salvio:2014soa}.

 \vskip0.2cm
Although in higher curvature gravity and its solutions \cite{Jacobson:1993xs}, a linear Einstein term was assumed to be always present in the action, the case of pure quadratic curvature theories has also been considered \cite{Deser,Kehagias:2015ata}.  In particular, as  was recently pointed out \cite{Kounnas:2014gda},  
the pure   ${ R}^2$ is interesting for the following reasons: it is the only pure quadratic theory that it is free of ghosts and scale invariant. The latter is a classical statements, which however is expected to be violated quantum mechanically leading to the emergence among others of the Einstein term. 
The pure $R^2$ theory is conformally equivalent to Einstein gravity with a minimally coupled scalar field and a cosmological constant. 

 \vskip0.2cm

In this paper we  discuss the spectrum of quadratic gravity theories, and in particular, the physical spectrum of the pure $R^2$ theory. 
Confirming that this theory is free of ghosts, 
we find that the spectrum of this theory nevertheless depends on the background: we show that
on a flat background the theory propagates only a scalar mode and no spin-2 field, whereas on a curved space it propagates a massless spin-2 graviton and a spin-0 scalar 
state. 
This result is in agreement with the observation that the pure $R^2$ theory is only conformally equivalent to Einstein gravity with a minimally coupled scalar field and a cosmological constant, when formulated on a curved background with $R\neq 0$. On the other hand, for a  background with $R=0$, this conformal transformation is singular.

 \vskip0.2cm

 So a sensible theory of gravity with a propagating spin-2 gravity field can only be obtained if one puts  the $R^2$ theory, which is relevant for the 
short distance physics in the UV,
 on a background, which still possesses
 some  curvature at long distances in the IR. Alternatively the theory starts to gravitate, if one adds to the UV $R^2$ theory  the standard Einstein term, which is responsible for the long
range gravity interactions in the IR. Hence there is an interesting  interplay between UV and IR physics in theories of quadratic gravity.

 \vskip0.2cm

In the paper we also discuss the Newtonian limit of these theories and we calculate  the deviation from Newton's law. Finally, we  will show how pure quadratic gravity theories may be obtained from a string-theory set up. In particular, string theory on a non-compact $CY_3$ of relatively small Euler number and large non-vanishing 4-cycles
may lead in a certain limit to a pure Riemann square  gravity.
We investigate in detail the case of a line bundle over $\mathbb{CP}^2$ and we show explicitly how  a pure Riemann square theory is obtained,  showing that  the associated ghost in this case is due to a truncation of the string effective action.

\vskip0.2cm

The structure of this paper is as follows: In section 2 we present the relation between the various quadratic curvature terms. In section 3 we investigate the physical spectrum of  quadratic actions by considering fluctuations on flat and curved backgrounds. In section 4 we discuss the Newtonian limit of these theories. In section 5 we explore the possibility for obtaining quadratic gravity actions as a particular limit of the string effective theory. We conclude  in 
section 6.

\vskip0.2cm

There is a large literature on the study of the so-called $f(R)$ theories
of gravity (see for instance \cite{DeFelice:2010aj} and references
therein).  More recently, the cosmological properties of $R^2$-like
theories and some of their black hole properties have been studied
in \cite{Cognola:2015uva,Cognola:2015wqa,Rinaldi:2014gha,Rinaldi:2014gua}.
The absence of propagating spin 2 modes around flat backgrounds
was also encountered in studies of three-dimensional massive
gravity \cite{Bergshoeff:2009aq,Bergshoeff:2009hq}.


\section{The Higher Curvature Action and Field equations}

The most general formulation of   scale invariant gravity is described  by an action  containing the following three terms being second order in the curvature tensor:
\begin{eqnarray}
S=\int_{\cal M} d^4 x\sqrt{-g}\Big{(}c_1 C_{\mu\nu\rho\sigma}C^{\mu\nu\rho\sigma}+c_2 R^2+c_3\hat R_{\mu\nu}^2\Big{)}. \label{W}
\end{eqnarray}
The first term with $C_{\mu\nu\rho\sigma}$ being the Weyl tensor is conformally invariant, whereas the $R^2$ term 
and the $\hat R_{\mu\nu}^2$
are only scale invariant.  $\hat R_{\mu\nu}=R_{\mu\nu}-{1\over 4}g_{\mu\nu}R$ is the traceless part of the Ricci tensor. All three couplings $c_i$ are dimensionless.

At the classical level one can reduce the number of independent quadratic curvature terms in the action (\ref{W})  from three to two. 
This is possible since
one can replace each of three quadratic terms by the other two plus the Gauss-Bonnet term
$GB$, which takes in four dimensions the following form:
\begin{equation}
GB=
  R_{\mu\nu\rho\sigma}R^{\mu\nu\rho\sigma}-4 R_{\mu\nu}R^{\mu\nu}
  +R^2.
  \end{equation}
The integral of the Gauss-Bonnet term over the space ${\cal M}$ is given by the topological Euler number $\chi({\cal M})$,
\begin{equation}
\chi(M)=\frac{1}{32\pi^2}\int_{\cal M} d^4x \sqrt{g}~GB+\int_{\partial M}d^4 x \sqrt{g} \Phi,
\end{equation}
where $\Phi$ is constructed from the second fundamental form of the boundary and the Riemann curvature and represents the boundary contribution. Clearly, for compact manifolds only the Gauss-Bonnet integral defines the Euler number,
 and it does not contribute to the classical equations of motion in this case. However for topologically non-trivial solutions like gravitational instantons, the GB term will lead to a non-trivial, topological
 contribution to the path integral of the higher curvature action.

Let us first eliminate the square of the Weyl tensor from the action (\ref{W}) using the following well-known relation:
\begin{eqnarray}
 C_{\mu\nu\rho\sigma}C^{\mu\nu\rho\sigma}&=&R_{\mu\nu\rho\sigma}R^{\mu\nu\rho\sigma}-2\hat R_{\mu\nu}^2-{1\over 6}R^2\nonumber\\
 &=&GB+2\hat R_{\mu\nu}^2-{1\over 6}R^2. \label{ident}
\end{eqnarray}
Using this substitution we can rewrite the action (\ref{W}) in the following form:
\begin{eqnarray}
S=\int_{\cal M} d^4 x\sqrt{-g}\Big{(}a(R_{\mu\nu}R^{\mu\nu}-{1\over 3}R^2) +bR^2+c ~GB
\Big{)}, \label{W2}
\end{eqnarray}
with the following relations among the coupling constants:
\begin{equation}
a=2c_1+c_3,\quad b=c_2+{c_3\over 12}, \quad c=c_1.
\end{equation}
We will use the action (\ref{W2})  in the next section  when we discuss the propagation modes of the theory.

Alternatively we can eliminate the $\hat R_{\mu\nu}^2$ term from (\ref{W}),  leading to:
\begin{eqnarray}
S=\int d^4 x\sqrt{-g}\Big{(}{a\over 2}  C_{\mu\nu\rho\sigma}C^{\mu\nu\rho\sigma}+b R^2+ g~GB\Big{)} \label{W1},
\end{eqnarray}
with $g=-c_3/2$.
This form of the action is convenient for deriving the classical field equations. Specifically 
the latter can be written as \cite{Kehagias:2015ata,pope1}
\begin{eqnarray}
W_{\mu\nu}=b J_{\mu\nu}- a B_{\mu\nu}=0, \label{JB}
\end{eqnarray}
where
\begin{equation}
J_{\mu\nu}=R\, R_{\mu\nu}-\frac{1}{4}R^2\, g_{\mu\nu}-\nabla_\mu\nabla_\nu R
+g_{\mu\nu}\nabla^2 R  \label{J},
\end{equation}
and  $B_{\mu\nu}$ is the Bach tensor:
\begin{eqnarray}
B_{\mu\nu}=\left(\nabla^\rho\nabla^\sigma+\frac{1}{2}R^{\rho\sigma}\right)C_{\mu\rho\nu\sigma}.
\end{eqnarray}
As recently discussed in \cite{Kehagias:2015ata},  these field equations 
are solved in particular by two distinct classes of solutions:

\vskip0.2cm
\noindent
(i) Spaces with vanishing scalar curvature:
\begin{equation}
R=0.
\end{equation}
This class of solutions of the $R^2$ theory contains in particular  non-Ricci flat spaces, $R_{\mu\nu}\neq 0$.  In section 3.1 we
will represent the pure $R^2$ theory in terms of an Einstein frame. These solutions will not be described by that theory.
 Therefore they deserve special attention, in particular in the description of the fluctuating modes around the scalar flat solutions
and the question about the absence of ghost when expanding around vacua with $R=0$.

\vskip0.2cm
\noindent
(ii) Einstein spaces satisfying:
\begin{eqnarray}
R_{\mu\nu}=3\lambda \, g_{\mu\nu} \label{dd},
\end{eqnarray}
where $\lambda$ is an {arbitrary} constant. In particular this class of solution contains (anti-)  de Sitter space.
We will also discuss below the propagating modes in such vacua.

\section{Physical modes and ghosts}

In this chapter we want to discuss the physical propagating modes of the higher curvature action (\ref{W}).
In particular we are interested in the question of the possible ghost modes in the pure $R^2$ theory. This can be done by expanding the action  (\ref{W}) in terms of
the fluctuations
around some particular curved or flat background with $\bar R\neq 0$ or $\bar R=0$ respectively. Alternatively, for the case of the 
 pure $R^2$ theory and for backgrounds with $R\neq 0$, one can investigate this problem by transforming the action to conventional Einstein gravity with a cosmological constant 
 \cite{Kounnas:2014gda,Kehagias:2015ata}.

 \subsection{Conformal transformation to Einstein frame\label{r2}}

Let us write the  pure $R^2$ action as
 \begin{eqnarray}
S=\int d^4 x\sqrt{-g}~  b\,  R^2.
\label{S1}
\end{eqnarray}
Here $b$ is the dimensionless coupling constant that can, as we will see, can be either positive or negative.
This action 
can equivalently be written as 
\begin{eqnarray}
S=\int d^4 x \sqrt{-g}\left(\Phi R-\frac{1}{4b} \Phi^2\right) .
\label{S4}
\end{eqnarray}
The dimension 2 scalar field $\Phi$ plays the role of a Lagrange multiplier and arises in this conformal (Jordan) frame without space-time derivatives. Through its equation of motion
$\Phi$ is proportional  to the background scalar curvature $\bar R$:
\begin{equation}
\Phi=2 b\bar R.
\end{equation}
Performing a conformal transformation 
\begin{equation}\label{weyl}
g_{\mu\nu}= {1\over 2}M_P^2\Phi^{-1}\tilde g_{\mu\nu}
\end{equation}
the action (\ref{S4}) can  be written as  
\begin{eqnarray}\label{einst}
S=\int d^4 x \sqrt{-\tilde g}\left(\frac{M_P^2}{2}\tilde R-\frac{3M_P^2}{4}{\partial_\mu \Phi \partial_\nu \Phi\over\Phi^2} -\frac{M_P^4}{16 b}\right) \label{E}.
\end{eqnarray}

 Needless to say, most of the manipulations so far, and those that will follow require the scalar curvature to be different from zero.  If
that is the case, we  have in particular:
 \vskip0.2cm
\noindent
(i) de Sitter backgrounds:
\begin{equation}
\bar R>0,\quad b>0.
\end{equation}
In this case $\Phi$ is positive and hence also $M_P^2>0$, if we require that the conformal transformation (\ref{weyl}) does not change the signature of the metric.
This class of solutions of the $R^2$ theory describes de Sitter like backgrounds with positive cosmological constant $\Lambda=M_P^4/16 b$ in
the Einstein frame. Both the signs of the Einstein term as well as of the scalar kinetic term in the Einstein action (\ref{einst}) are such that the spin-2 graviton as well as the scalar
field $\Phi$ are physical, ghost-free degrees of freedom. Hence gravity acts as an attractive force.  This will be further discussed in the Newtonian limit in section \ref{Newt}.

 \vskip0.2cm
\noindent
(ii) anti-de Sitter backgrounds:
\begin{equation}
\bar R<0,\quad {b}<0.
\end{equation}
 $\Phi$ and $M_P^2>0$ are again positive. Now
this class of solutions of the $R^2$ theory describes anti-de Sitter like backgrounds with negative cosmological constant $\Lambda=M_P^4/16 {b}$ in
the Einstein frame. Again the spin two graviton as well as the scalar
field $\Phi$ are physical, ghost-free degrees of freedom.

In addition  to these two cases there are also two further choices $R$ and $b$ leading to unphysical, ghost-like propagating modes:
 \vskip0.2cm
\noindent
(iii,iv) de Sitter backgrounds:
\begin{equation}
\bar R<0,\quad {b}>0\, \quad{\rm and}\quad \bar R>0,\quad {b}<0.
\end{equation}
In these two cases $\Phi$ is negative. Hence the scalar mode in (\ref{einst}) is ghost-like and the sign of the Einstein term is such that gravity acts
as a repulsive force.

This analysis shows that with the correct identifications of the parameters, the pure $R^2$ is indeed ghost-free around (anti-) de Sitter backgrounds, and the propagating modes correspond to two spin-2 graviton degrees of freedom plus
one scalar with physical kinetic energy.
However the above simple analysis relies on the existence of the conformal transformation (\ref{weyl}), that necessitates backgrounds with $R\neq 0$ throughout
space-time. 
For backgrounds with $\bar R=0$  this transformation is singular.
 As argued in \cite{Kehagias:2015ata}, this limit  is similar to the tensionless string limit of string theory in
six dimensions.
In any case,  for flat background with $\bar R=0$ the conclusion about the ghost-freedom of $R^2$ gravity cannot be drawn immediately, and it is more reliable to directly analyze the
propagating modes of the higher curvature action.

 \subsection{Propagating modes in higher curvature actions}

Let us write the action in the following form, where we also include the standard Einstein term plus a cosmological constant:
\begin{eqnarray}
S=\int d^4 x\sqrt{-g}\Big{(}a(R_{\mu\nu}R^{\mu\nu}-{1\over 3}R^2) +bR^2+\kappa^2 R+\lambda\Big{)}. \label{W4}
\end{eqnarray}
The first two terms with coefficients $a$ and $b$ are the same as in the action (\ref{W2}).

Now we study  the propagating modes corresponding to this action. For this, we analyze  the poles in the propagators 
generated by its quadratic part. This was already done in \cite{Julve:1978xn}, and we are just stating the results of this paper.
Specifically, there are three kinds of propagating modes:

\vskip0.5cm
\noindent
(i) A massless spin 2 graviton: this mode is independent of $a,b,\kappa^2$. It is the standard massless spin 2 graviton.

\vskip0.2cm
\noindent
(ii)  A massive spin two particle with mass $\kappa^2/(-a)$. It is related to the Weyl$^2$ term in the action. In fact, this spin two state is either a tachyonic ($a>0$), or a massive ghost ($a<0$). So one should  get rid of it.
\vskip0.2cm
\noindent
(iii)  A massive scalar with mass proportional to $\kappa^2/6b$. It is related to the $R^2$ term in the action.

\vskip0.5cm
 The pure $R^2$ is recovered in the  $a,\kappa^2,\lambda\rightarrow 0$ limit. But this limit is rather delicate  and has to be taken with  care.
So we will
refine the above discussion and  will set $a=\lambda=0$, {\it i.e.} the action is
\begin{eqnarray}
S=\int d^4 x\sqrt{-g}\Big{(}bR^2+\kappa^2 R\Big{)}. \label{W11}
\end{eqnarray}
This action includes the Starobinski model for $\kappa\neq0$, $b\neq 0$ as well as the pure $R^2$ theory in case $\kappa=0$.

\subsubsection{Flat spaces solutions \label{fb}}

First we expand the action (\ref{W11}) around  Minkowski space  
$g_{\mu\nu}=\eta_{\mu\nu}+h_{\mu\nu}$.
The scalar fluctuation is denoted by $\varphi={h_\mu}^\mu$, 
and  the action (\ref{W11}) takes the form
\begin{eqnarray}
S_0=\int d^4 x\sqrt{-g}\,\biggl\{ {\kappa^2\over 2}
\left(\frac{1}{2}h_{\mu\nu} \Box h^{\mu\nu}-\frac{1}{2}\varphi\Box \varphi - D_\mu \varphi D_\nu h^{\mu\nu}
+ D^\mu h_{\mu\rho} D_\nu h^{\nu\rho}\right)
+b(\partial^\mu\partial^\nu h_{\mu\nu}-\square\varphi)^2\biggr\}.   \label{W12}
\end{eqnarray}
As usual, we can decomposed the symmetric tensor $h_{\mu\nu}$ as 
\begin{eqnarray}
h_{\mu\nu}=h_{\mu\nu}^\perp+\partial_\mu a_\nu^\perp+
\partial_\nu a_\mu^\perp+(\partial_\mu\partial_\nu-
\frac{1}{4}\eta_{\mu\nu}\Box)a+\frac{1}{4}\eta_{\mu\nu}\varphi, \label{dec}
\end{eqnarray}
where $h_{\mu\nu}^\perp$ is transverse traceless
\begin{eqnarray}
\partial^\mu h_{\mu\nu}^\perp=\eta^{\mu\nu}h_{\mu\nu}^\perp=0,
\end{eqnarray}
and $a_\mu^\perp$ is transverse, {\it i.e.}, divergenceless
\begin{eqnarray}
\partial^\mu a_\mu^\perp=0.
\end{eqnarray}
Gauge transformations are the infinitesimal diffeomorphisms $x^\mu\to x^\mu+\xi^\mu(x)$ under which the metric transforms as:
\begin{eqnarray}
h_{\mu\nu}\to h_{\mu\nu}+\partial_\mu\xi_\nu+\partial_\mu\xi_\nu .\label{gt}
\end{eqnarray}
Note that we may also decompose $\xi_\mu$ in transverse and longitudinal 
parts as 
\begin{eqnarray}
\xi_\mu=\xi_\mu^\perp+\partial_\mu\xi
\end{eqnarray}
with 
\begin{eqnarray}
\partial^\mu\xi_\mu^\perp=0, ~~~\Box\xi=\partial^\mu\xi_\mu.
\end{eqnarray}
With this decomposition, under the gauge transformation (\ref{gt}), we get that
\begin{eqnarray}
&&h_{\mu\nu}^\perp\to h_{\mu\nu}^\perp,\\
&& a_\mu^\perp\to a_\mu^\perp+\xi_\mu^\perp,\\
&&a\to a+2\xi ,\\
&&\varphi\to \varphi+2\Box \xi. \label{tr1}
\end{eqnarray}
Therefore, the field 
\begin{eqnarray}
\Phi=\varphi-\Box a
\end{eqnarray}
is invariant under (\ref{gt}), { i.e.}
\begin{eqnarray}
\Phi\to \Phi. \label{trf}
\end{eqnarray}
Then, it is easy to verify that 
\begin{eqnarray}
\partial^\mu\partial^\nu h_{\mu\nu}-\Box \varphi=\frac{3}{4}\left(\Box^2 a-\Box \varphi\right)=-\frac{3}{4}\Box \Phi
\end{eqnarray}
and, hence the quadratic action (\ref{W12}) becomes:
\begin{eqnarray}
S_0=\int d^4 x\sqrt{-g}\bigg\{{\kappa^2\over 4}\left(h_{\mu\nu}^\perp\square {h^\perp}^{\mu\nu}+\partial_\mu\Phi\partial^\mu\Phi\right)+{9b\over 16}(\square\Phi)^2\bigg\}.   \label{W112}
\end{eqnarray}
It should be noted that  that  the Fourier modes $\tilde h_{\mu\nu}$ of $h_{\mu\nu}$  have a  decomposition similar to (\ref{dec}).
In particular we have that 
\begin{eqnarray}
&&\tilde h_{\mu\nu}^\perp=P_{\mu\nu,\rho\sigma}^{(2)} \tilde h_{\rho\sigma}\\
&&\tilde \Phi =P_{\mu\nu,\rho\sigma}^{(0)} \tilde \Phi
\end{eqnarray}
where $P_{\mu\nu,\rho\sigma}^{(2)}, ~P_{\mu\nu,\rho\sigma}^{(0)}$ are the projectors for the spin-2 and spin-0 parts of $h_{\mu\nu}$ in momentum space. They are explicitly given by
\begin{eqnarray}
&&P_{\mu\nu,\rho\sigma}^{(2)}=\frac{1}{2}\left(\theta_{\mu\rho}\theta_{\nu\sigma}+\theta_{\nu\rho}\theta_{\mu\sigma} \right)-
\frac{1}{3}\theta_{\mu\nu}\theta_{\rho\sigma},\nonumber \\
&&P_{\mu\nu,\rho\sigma}^{(0)}=\frac{1}{3}\theta_{\mu\nu}\theta_{\rho\sigma}, ~~~\theta_{\mu\nu}=\eta_{\mu\nu}-\frac{q_\mu q_\nu}{q^2}.
\end{eqnarray}
In the Starobinski case around flat space there is then a massless spin-2 fluctuation with propagator
\begin{equation}
\Delta_{\mu\nu,\rho\sigma}^{(2)}=-{2\over \kappa^2}{1\over q^2}P_{\mu\nu,\rho\sigma}^{(2)}. \label{s-22}
\end{equation}
It disappears in the limit $\kappa=0$, so in the  pure $R^2$ theory around flat space there is no propagating spin-2 graviton.

Next, the spin zero propagator reads
\begin{eqnarray}\label{spinzeroprop}
\Delta_{\mu\nu,\rho\sigma}^{(0)}&=&-{1\over 6b}{1\over q^2(q^2+\kappa^2/6b)}P_{\mu\nu,\rho\sigma}^{(0)}\nonumber\\
&=&-{1\over  \kappa^2}\biggl({1\over q^2+\kappa^2/6b-i\epsilon}-{1\over q^2+i\epsilon}\biggr)P_{\mu\nu,\rho\sigma}^{(0)}, \label{s0}
\end{eqnarray}
where we have also implemented an $i\epsilon$ prescription for  the spin-zero propagator. 
In the Starobinski case with $\kappa,b\neq0$,  $\Delta_{\mu\nu,\rho\sigma}^{(0)}$ describes two scalar modes; the first term in (\ref{spinzeroprop}) corresponds to a massive spin zero particle with mass 
\begin{equation}
m^2={\kappa^2\over 6b}.
\end{equation}
It describes the massive scalar excitation around the flat Starobinski vacuum. The second term in (\ref{spinzeroprop}) describes  a massless 
spin zero ghost. 
Note that the propagators for the normal massive state and the ghost have opposite imaginary parts. Choosing the same prescription for the normal and the ghost state would result in a violation of the optical theorem and a corresponding violation of unitarity. However, with the above $+i\epsilon$ prescription for the ghost, unitarity is maintained at the cost of 
propagation of negative energy forward
in time \cite{Cline:2003gs}.

In should be noted that the ghost state encountered above, is the standard ghost emerging also in general relativity. Indeed, by setting $b=0$, i.e in the  general relativity limit, we see  that $\Phi$ is gauge invariant and more surprisingly it is a ghost. Thus, it seems that general relativity propagates an additional state besides the two helicity $\pm 2$ graviton states. However this ghost is harmless as it is not   a propagating physical  mode. To see this, 
let us recall that 
 $\Phi$ is written in terms of $h_{\mu\nu}$ as 
\begin{eqnarray}
\Phi=P^{\mu\nu}h_{\mu\nu}
\end{eqnarray}
where 
\begin{eqnarray}
P^{\mu\nu}=\eta^{\mu\nu}-\frac{4}{3}\Box^{-1}
\left(\partial^\mu\partial^\nu-\frac{1}{4}\eta^{\mu\nu}\Box\right).
\end{eqnarray} 
As a result, $\Phi$ is non-local in time so that the initial data
 for the metric perturbations $h_{\mu\nu}$  at  a given time, are not enough to determine $\Phi$ at a later  time. Therefore,  there is no extra degree of freedom in general relativity as a  naive counting
 indicates. This can also  be  seen from the fact that although $h_{\mu\nu}^\perp$ and $\Phi$ are gauge invariant (\ref{tr1},\ref{trf}), there is still a residual gauge symmetry. Indeed, 
we may still transform $h_{\mu\nu}^\perp$ and $\Phi$
as 
\begin{eqnarray}
 &&h_{\mu\nu}^\perp\to h_{\mu\nu}^\perp+D_\mu\xi_\nu+D_\nu\xi_\mu,\label{res1}\\
 && \Phi\to \Phi+2  D^\mu k_\mu, \label{res2}
 \end{eqnarray} 
 provided $\xi_\mu,~k_\mu$ satisfy
 \begin{eqnarray}
 D^\mu\xi_\mu=0,~~~\left(\Box+\frac{R}{4}\right) \xi_\mu=0, ~~~ 
 D_\mu k_\nu+4 D_\nu k_\mu=\frac{1}{2}g_{\mu\nu}D^\sigma k_\sigma, 
 \end{eqnarray}
$k_\mu$ being  just conformal Killing vectors. Therefore, $\xi_\mu$ eliminates four of the six components of $h_{\mu\nu}^\perp$ leaving two propagating components which correspond to the $\pm 2$ helicity states of the graviton, whereas, the divergence of $k_\mu$ eliminates the would-be spin-0 ghost state $\Phi$. 

As a result, in the Starobinsky theory, the massless ghost state appearing in the propagator (\ref{s0}) is nothing else than the usual
harmless non-propagating state already encounter in general relativity, and can be eliminated either due to its non-local relation to the metric perturbation or by gauge symmetry arguments. Alternatively, we may write the spin-0 part of (\ref{W112}) as 

\begin{eqnarray}
S_0&=& \int d^4 x\left({3\kappa^2\over 32}\partial_\mu\Phi\partial^\mu\Phi+{9b\over 16}(\square\Phi)^2 +\Phi T+\cdots\right) \nonumber \\
&=& 
 \int d^4 x\left({3\kappa^2\over 32}\partial_\mu\Phi\partial^\mu\Phi+\Psi \Box \Phi -{4\over 9b}\Psi^2 +\Phi T+\cdots\right),\label{W1112}
\end{eqnarray} 
where the coupling of $\Phi$ to the trace $ T_\mu^\mu=8  T$ of the energy-momentum tensor has also been included.
Then the field equations for $\Psi$ and $\Phi$ are 
\begin{eqnarray}
&&-\frac{3\kappa^2}{16}\Box \Phi+\Box \Psi=-T,\\
&&\Box\Phi=\frac{8}{9b}\Psi,
\end{eqnarray}
or equivalently:
\begin{eqnarray}
&&-\frac{\kappa^2}{6b}\Psi+\Box \Psi=-T,\\
&&\Box\Phi=\frac{8}{9b}\Psi.
\end{eqnarray}
 Again, we see  that the theory  propagates only a massive scalar  degree of freedom $\Psi=9b/8  \, \Box \Phi$ with mass $m_\Psi^2=\kappa^2/6b$, whereas $\Phi$ is a not propagating field.



\vskip0.2cm

Next we discuss the scalar modes in the  pure $R^2$ theory around flat space. In the limit $\kappa=0$, the scalar propagator in (\ref{spinzeroprop}) becomes
\begin{eqnarray}
\Delta_{\mu\nu,\rho\sigma}^{(0)}=-{1\over 6b}{1\over q^4}P_{\mu\nu,\rho\sigma}^{(0)}
.
\end{eqnarray}
It describes a massless dipole, {\it i.e.}  a pair of massless scalars; however to make the propagator well-defined an $i\epsilon$ description should be provided. 
We will follow a slightly different approach  to analyze the spectrum.
For the pure $R^2$ theory, the action (\ref{W1112}) is written as  

\begin{eqnarray}
S_0= \int d^4 x\sqrt{-g}\biggl({9b\over 16}(\square\Phi)^2+\Phi T
+\cdots\biggr) = 
 \int d^4 x\sqrt{-g}\left(\Psi \Box \Phi -{4\over 9b}\Psi^2 +\Phi T+\cdots \right)\label{W1113}
\end{eqnarray} 
 and the equations for $\Phi,\Psi$ in the presence of a source are now
 \begin{eqnarray}
&&\Box \Psi=-T,\\
&&\Box\Phi=\frac{8}{9b}\Psi.
\end{eqnarray}
Thus, $\Psi$ is a normal propagating  massless degree of freedom, whereas, $\Phi$ is not propagating as it is a non-local function of the metric perturbations or 
differently put, it can be fixed by the residual gauge symmetry (\ref{res2}).  As a result, the pure $R^2$ theory propagates just a single spin-0 state around flat Minkowski
 background. In particular, no tensor mode can be excited for this theory on flat spacetime.  So the pure $R^2$ theory on flat spaces does not gravitate.

\vskip0.2cm

Let us also note at this point that for the action (\ref{W4}), although the propagator of the spin-0 part of  $h_{\mu\nu}$ is still given by 
(\ref{s0}), the spin-2 propagator turns out to be 
\begin{equation}
\Delta_{\mu\nu,\rho\sigma}^{(2)}=-{2\over q^2(q^2a+\kappa^2)}P_{\mu\nu,\rho\sigma}^{(2)}. \label{ww12}
\end{equation}
Therefore, it reduces to (\ref{s-22}) for $a=0$ as expected and there is an additional pole at $q^2=-\kappa^2/a$ corresponding to a massive spin-2 ghost.  

\subsubsection{Curved spaces solutions \label{cb}}

We now consider perturbations around the de Sitter background 
\begin{eqnarray}
g_{\mu\nu}=\bar{g}_{\mu\nu}+h_{\mu\nu},\quad {\rm with}\quad \bar{R}_{\mu\nu}=\lambda \bar{g}_{\mu\nu},\quad\lambda={\bar R\over 4}
.
\end{eqnarray}
Then we find to second order: 
\begin{align}
S_2=b\int d^4 x&\left\{ \left[ D_\mu D_\nu h^{\mu\nu}-\Box h-\frac{1}{4}\bar R \varphi\right]^2 -\bar R \left(-\frac{1}{2}h_{\mu\nu} \Box h^{\mu\nu}+\frac{1}{2}\varphi\Box \varphi +D_\mu \varphi D_\nu h^{\mu\nu}
\right.\right.\nonumber \\ 
&\left.\left. \phantom{\frac{1}{4}}
-D^\mu h_{\mu\rho} D_\nu h^{\nu\rho}\right)+\frac{\bar{R}^2}{6}\left[h_{\mu\nu}h^{\mu\nu}+\frac{1}{4} \varphi^2\right]
\right\},
\end{align}
where $D_\mu$ is the covariant derivative with respect to the de Sitter metric $\bar{g}_{\mu\nu}$. We may again decompose 
$h_{\mu\nu}$ as 
\begin{eqnarray}
h_{\mu\nu}=h_{\mu\nu}^\perp+D_\mu a_\nu^\perp+
D_\nu a_\mu^\perp+(D_\mu D_\nu-
\frac{1}{4}\eta_{\mu\nu}\Box)a+\frac{1}{4}\bar{g}_{\mu\nu}\varphi,
\end{eqnarray}
where 
\begin{eqnarray}
D^\mu h_{\mu\nu}^\perp=\eta^{\mu\nu}h_{\mu\nu}^\perp=D^\mu a_\mu^\perp=0,
\end{eqnarray}
so that, the quadratic action turns out to be
\begin{align}
S_2=b\int d^4 x&\left\{ \frac{9}{16}\left[ \Box\Phi+\frac{\bar R}{3} \Phi\right]^2 + \bar R \left[\frac{1}{2}h_{{\mu\nu}}^\perp \left( \Box -\frac{\bar{R}}{6}\right)h^{{\mu\nu}^\perp }
-\frac{3}{16}\Phi\left(\Box +\frac{\bar{R}}{3}\right)\Phi\right]
\right\} \nonumber \\
&=b\int d^4 x\left\{\frac{\bar R}{2}h_{{\mu\nu}}^\perp \left( \Box -\frac{\bar{R}}{6}\right)h^{{\mu\nu}^\perp }+\frac{9}{16}\Phi\left(
\Box^2+\frac{\bar R}{3}\Box\right)\Phi.
\right\}
 \label{ac0}
\end{align}
This structure of the quadratic action can in fact be obtained by recalling that (with $\delta_1,\delta_2$ first order and second order variations respectively) 
\begin{align}
S_2&=b\int d^4 x\sqrt{-g}\Big{\{}(\delta_1 R)^2 \sqrt{-g} +2  \bar{R} (\delta_1R)(\delta_1\sqrt{-g})+\bar R ^2(\delta_2 \sqrt{-g})+2 \bar R (\delta_2 R)
\sqrt{-g}\Big{\}}\nonumber \\
&=b\int d^4 x\sqrt{-g}\Big{\{}(\delta_1 R)^2 \sqrt{-g}+ 
2\bar R \delta_2(R \sqrt{-g})-\bar R ^2(\delta_2\sqrt{-g})\Big{\}}. \label{acc}
\end{align}
The last two terms of the above action,  giving  rise to the last two terms in (\ref{ac0}), are exactly the quadratic part
$\delta_2 S_{EH}$ of the Einstein-Hilbert action 
\begin{eqnarray}
S_{EH}=\int d^4 x \sqrt{-g}\frac{M_P^2}{2}\biggl(R-2\Lambda\biggr)
\end{eqnarray}
with 
Planck mass $M_P$ and  a cosmological constant given by 
\begin{eqnarray}
\Lambda=\frac{ M_P^4}{16 {b}}
\end{eqnarray}
In addition, the first term in (\ref{acc}) gives rise to the first term in  (\ref{ac0}). This is a higher derivative term that gives rise to the  kinetic energy of $\Phi$.  In other words: pure $R^2$ theory on a de Sitter background propagates a massless spin-2 graviton and an additional massless spin-0 scalar excitation.

So there is no ghost in the pure $R^2$ in a background with constant
scalar curvature  $R\neq 0$.
 We  see, that going from the  flat space background to the (anti)-de Sitter background, the situation with respect to the gravity spectrum improves substantially.
This situation is similar to higher spin theories of the Vasiliev type (also a kind of $M_P\rightarrow 0$ limit of gravity), 
which are  only consistent in (anti)-de Sitter background but not in flat space.
The fact that the curvature of space 
plays a stabilizing role of the UV  theory in the IR, by providing propagating gravitons
and also a good large energy behaviour, is indeed remarkable.


Summarizing this section, we stress again that pure $R^2$ theory propagates just a single scalar degree of freedom  on flat spacetime and no spin-2 tensor mode. The latter can be recovered by IR regulating  the  theory by considering non-flat backgrounds (de Sitter or anti-de Sitter)  or by adding the Einstein term as in section (\ref{fb}). Both ways introduce back the graviton spin-2 state,  while keeping the massless spin-0 state.

\section{Newtonian Limit \label{Newt}}

The equations of motions  in the $R^2$ theory for a test particle following the trajectory 
$x^\mu(s)$  with $u^\mu=dx^\mu/ds$ are as usual 
\begin{eqnarray}
 \frac{du^\mu}{ds}+\Gamma^\mu_{\nu\rho}u^\mu u^\nu=0. \label{geo}
 \end{eqnarray} 
We may express the theory in the Einstein frame by using  the 
conformally transformed  metric $\bar{g}_{\mu\nu}=g^{E}_{\mu\nu}$ 
for $R(g)\neq 0$ 
\begin{eqnarray}
\bar g _{\mu\nu}=\frac{8\mu}{R} g_{\mu\nu}.
\end{eqnarray}
Therefore, in the Einstein frame, the geodesic equation (\ref{geo}) followed by a particle will be
\begin{eqnarray}
\frac{du^\mu}{ds}+\Gamma^\mu_{\nu\rho}u^\mu u^\nu=f^\mu, \label{geo1}
\end{eqnarray}
where 
\begin{eqnarray}
f^\mu=\frac{1}{2R}\left(\delta^{\mu}_\nu \partial_\rho R+\delta^\mu_\rho\partial_\nu R-g_{\nu\rho} \partial^\mu R\right)u^\nu u^\rho.
\end{eqnarray}
Let us now calculate the Newtonian force between two massive scalars $\phi$ of mass $m$ in quadratic gravity \cite{Stelle}. The Newtonian potential will be
\begin{eqnarray}
V=-\frac{1}{4m^2}\frac{1}{(2\pi)^3}\int d^3 \vec{k} {\cal M}_{nr} e^{-i\vec{k}\cdot \vec{r}}, \label{V}
\end{eqnarray} 
where ${\cal M}_{nr}$ is the non-relativistic amplitude for the $\phi+\phi\to \phi+\phi$ process. The interaction Lagrangian is 
\begin{eqnarray}
{\cal L}=\frac{1}{2}h^{\mu\nu} \tau_{\mu\nu},
\end{eqnarray}
where 
\begin{eqnarray}
\tau_{\mu\nu}=\partial_\mu\phi \partial_\nu \phi -\frac{1}{2}\eta_{\mu\nu}\left((\partial\phi)^2+m^2 \phi^2\right).
\end{eqnarray}
Then with incoming momenta $p_1,p_2$ and outgoing $p_3,p_4$, 
as in Fig. 1, 
\begin{figure}[H]
    \centering
    \includegraphics[width=0.3\textwidth]{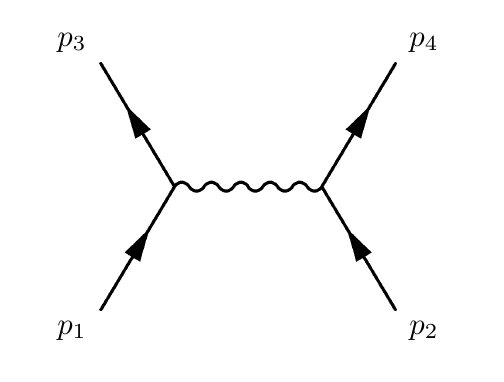}
    \caption{Digram for the calculation of the  Newtonian potential} \label{fig1}
\end{figure}
we have that  \cite{don}
\begin{eqnarray}
{\cal M}=\tilde{\tau}^{\mu\nu}(p_1,p_2)\Delta_{\mu\nu,\rho\sigma}
\tilde{\tau}^{\rho\sigma}(p_3,p_4), \label{amp}
\end{eqnarray}
where 
\begin{eqnarray}
\tilde{\tau}_{\mu\nu}(p_i,p_j)=\frac{1}{2}\left(p_{i\mu}p_{j\nu}+p_{j\mu}p_{i\nu}-\eta_{\mu\nu}(p_i\cdot p_j+m^2)\right), 
\end{eqnarray}
and $\Delta_{\mu\nu,\rho\sigma}$ is the graviton propagator. 
\vskip.2in

\noindent
{\bf Newtonial potential for the ${\mathbf{R+R^2+Weyl^2}}$ theory:} 
It is straightforward to calculate the Newtonian potential 
for the general action 
 \begin{eqnarray}
S=\int d^4 \sqrt{-g}\left(\kappa^2R+b\, R^2+\frac{a}{2}\, C_{\mu\nu\rho\sigma}C^{\mu\nu\rho\sigma}\right).
\label{S3}
\end{eqnarray}
In this case, by using Eqs.(\ref{s0}) and (\ref{ww12}),  we find that the amplitude  (\ref{amp}) is given by 
\begin{align}
{\cal M}&=\frac{1}{k^2(ak^2+\kappa^2)}
\left\{(p_1\cdot p_2)(p_3\cdot p_4)+(p_1\cdot p_3)(p_2\cdot p_4)
-(p_1\cdot p_2)(p_3\cdot p_4-m^2)\nonumber\right.\\
&\left.
 +2(p_1\cdot p_2-m^2)(p_3\cdot p_4-m^2)-\frac{2}{3}\Big{[}( p_1\cdot p_2 -2m^2)(p_3\cdot p_4-2m^2)\Big{]}\right\}\nonumber \\ &
 +\frac{1}{12} \frac{1}{k^2(6bk^2+\kappa^2)}\left(2m^2-p_1\cdot p_3\right)
\left(2m^2-p_3\cdot p_4\right).
\end{align}
Then, in 
the non-relativistic limit we get 
\begin{align}
{\cal M}_{nr}&=-\frac{4}{3}\frac{m^4}{|\vec{k}|^2(a\,|\vec{k}|^2+\kappa^2)}
+\frac{1}{3}\frac{m^4}{|\vec{k}|^2(6b \,|\vec{k}|^2-\kappa^2)}.
\label{mm}
\end{align}
It is clear that the first term in the rhs of (\ref{mm}) is due to the massless spin-2 graviton, where the second term in due to the spin-0 state. 
We can express (\ref{mm}) as
\begin{eqnarray}
{\cal M}_{nr}=-\frac{m^4}{\kappa^2}\left(\frac{1}{|\vec{k}|^2} -\frac{4}{3}\frac{a}{a|\vec{k}|^2+\kappa^2}-
\frac{2b}{6b\,|\vec{k}|^2-\kappa^2}\right).
\label{mm1}
\end{eqnarray}
After Fourier transforming and using (\ref{V}), we find that the Newtonian potential  for the theory (\ref{S3}) is given by
\begin{eqnarray}
V=-\frac{G m}{r}+\frac{4Gm}{3}\frac{e^{-\alpha M_p r}}{r}-\frac{Gm}{3}\frac{e^{-\beta M_p r}}{r}\label{V3}
\end{eqnarray}
where 
\begin{eqnarray}
\alpha=1/\sqrt{2a}, ~~~\beta=1/2\sqrt{3b},
 ~~~\kappa^2=1/16\pi G=M_p^2/2.
\end{eqnarray}
Interestingly, for $r\gg r_0$ where $r_0=1/min(\alpha,\beta) M_p$, we get the usual Newton law. However, for $r\ll r_0$ we get that 
\begin{eqnarray}
V\approx V_0-\frac{1}{48\pi}(\beta^2-4\alpha^2)m r,
\end{eqnarray}
where $V_0=Gm(\beta-4\alpha)M_p/3$. We see that the potential is finite at $r=0$ and it is confining for 

\be
\beta^2<4\alpha^2,
\ee
(it is also confining for $\beta=2\alpha$). 
Therefore, for $\beta^2\leq4\alpha^2$, the potential is a monotonic function of r, whereas, for $\beta^2>4\alpha^2$ it necessarily develops a  minimum as shown  in Fig. \ref{fig2}. 
\begin{figure}[H]
    \centering
    \includegraphics[width=0.5\textwidth]{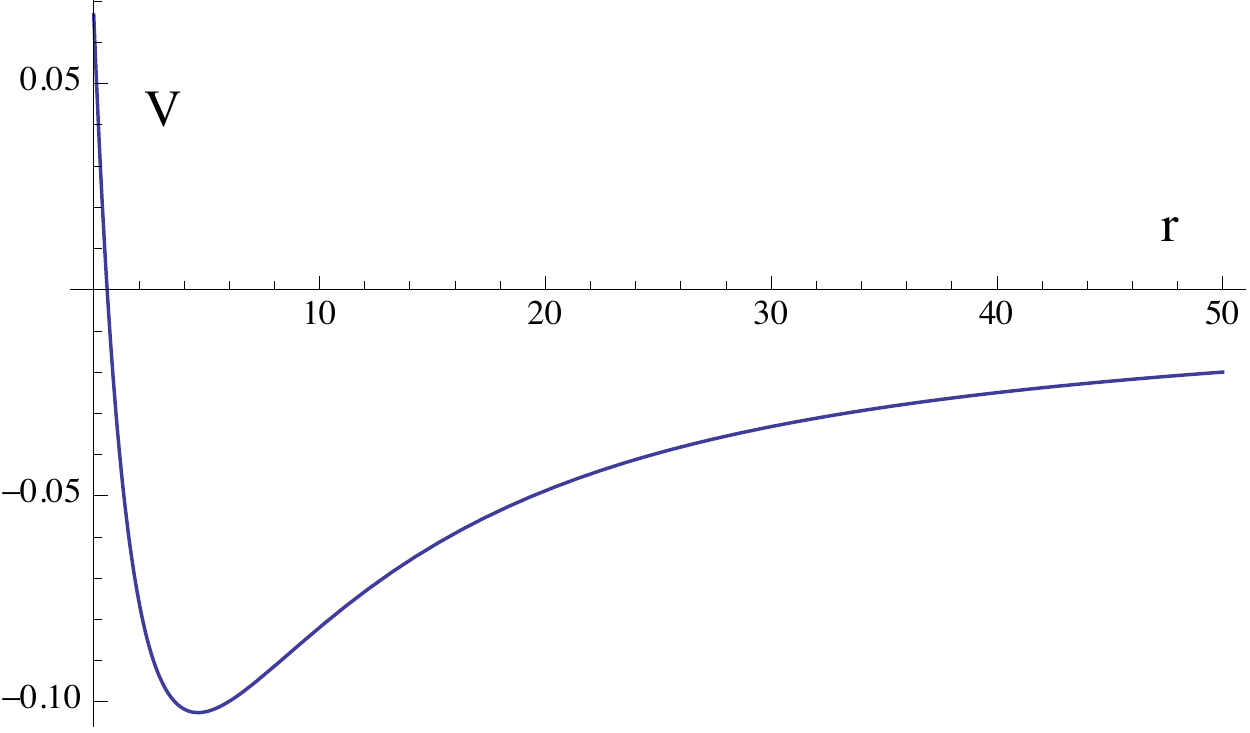}
    \caption{Newtonian potential for $\alpha=.2,\beta=1$.} \label{fig2}
\end{figure}

We may also employ the general expression (\ref{mm1}) to find the potential in some particular limits. 
\vskip.2in

\noindent
{\bf Newtonial potential for the ${\mathbf{R+R^2}}$ theory:} By setting $a=0$ in (\ref{mm1}) we get that 
\begin{eqnarray}
{\cal M}_{nr}=-\frac{m^4}{\kappa^2}\left(\frac{1}{|\vec{k}|^2} -
\frac{2b}{6b\,|\vec{k}|^2-\kappa^2}\right),
\label{mm11}
\end{eqnarray}
and the  Newtonian potential  for the $R+R^2$ theory turns out to be
\begin{eqnarray}
V=-\frac{G m}{r}-\frac{Gm}{3}\frac{e^{-\beta M_p r}}{r}.\label{V31}
\end{eqnarray}
Therefore, the usual Newtonian potential acquires an additional   Yukawa contribution from the massive spin-0 state.
\vskip.2in

\noindent
{\bf Newtonial potential for the ${\mathbf{R^2}}$ theory:} 
In the  $\alpha=0,~\kappa=0$ limit, the first term in the rhs of  (\ref{mm}) is absent and the  we get that 
\begin{eqnarray}
{\cal M}_{nr}=\frac{1}{18 {b}}\frac{m^4}{|\vec{k}|^4}.
\end{eqnarray}
The Newtonian potential is then given by
\begin{eqnarray}
V=-\frac{1}{72 {b}}m^2 \frac{1}{(2\pi)^3}\int d^3 \vec{k}\frac{ e^{-i\vec{k}\cdot \vec{r}}}{|\vec{k}|^4}
\end{eqnarray}
which, after Fourier transforming the generalized function 
$1/|\vec{k}|^4$, we find that
\begin{eqnarray}
V=\frac{1}{9\cdot 2^6\, \pi {b} }m^2 r,
\end{eqnarray}
{\it i.e.}  it is a confining potential without a Newtonian tail.

  \section{The quadratic action in String and M-Theory}

In string and M-theory higher curvature actions appear naturally in the effective field theory description of the light modes.
They are also  generated by compactification and decoupling of the massive Kaluza-Klein states.  The internal manifold
can in fact be chosen to be compact or non-compact.  We will explore the origin of higher curvature terms using
various compactifications.

\subsection{Compact internal spaces}
 We start with a  generic higher curvature expansion in the 11-dimensional M-theory:
\begin{equation}\label{M}
S_M=\int d^{11} x\sqrt{-G_{11}}{1\over l_{11}^9}\biggl(\sum_{n=1}^\infty l_{11}^{2(n-1)}{\cal R}_{11}^n\biggr).
\end{equation}
Here $l_{11}$ denotes the 11-dimensional Planck length and the higher curvature terms ${\cal R}_{11}^n$ comprise all possible n$^{\rm th}$ powers of 11-dimensional Riemann tensors,
Ricci tensors and Ricci scalars. Actually in 11-dimensional supergravity some of these terms are absent due to supersymmetry conditions.

In the first step,  let us reduce the 11-dimensional M-theory on a circle of radius $R_{11}$, deriving in this way the 10-dimensional type IIA theory. 
This is done by the well-known ansatz
\begin{equation}
l_{11}^2=\alpha',\quad R_{11}=(\alpha')^{1/2},
\end{equation}
where $e^\phi=g_s$ is the string coupling constant $(\alpha')^{-2}$ the string tension. After further rescaling of the 10-dimensional metric,
$G_{10}=e^{-{2\phi\over 3}}g_{10}$, the generic 10-dimensional type II action takes the form:
\begin{equation}
S_{10,IIA}=\int d^{10} x\sqrt{-g_{10}}\biggl(( \alpha')^{-4}\sum_{n=1}^\infty (\alpha')^{(n-1)}e^{{2\over 3}(n-4)\phi}{\cal R}_{10}^n\biggr)\, 
\end{equation}
Actually one has to remark that the terms in the 11-dimensional action, written down so far, are not the only terms in the M-theory $R^n$ expansion after compactification.
Namely, upon compactification on a circle, there are additional M-theory loop effects, which contain additional powers of $(l_{11}/R_{11})^l$, where l is basically the loop order
in M-theory.
They vanish in the limit $R_{11}\rightarrow \infty$, {\it i.e.} they are invisible in the 11-dimensional decompactification limit.
In the 10-dimensional ${\cal R}_{10}^n$ expansion they lead to terms which scale like $(\alpha')^{n-5}$, however due to the additional $R_{11}^{-1}$, they
scale differently with respect to the string coupling constant $e^\phi$.


Furthermore,  the 11-dimensional terms written in (\ref{M}), correspond not only to tree-level 10-dimensional IIA string theory, but contain also
loop contributions in the string genus expansion.
The precise relation is $2n-8 = 3(h-1)$, where $n$ is the power of ${\cal R}^n_{10}$ and $h$ is the  genus, i.e. the loop order in IIA.
Therefore an arbitrary value of $n$ seems to  give rise to  non-integer values of $h$, which are not easily exmmmmplained  in string perturbation theory. However, certain terms  are not present in 10-dimensional IIA action, like ${\cal R}_{10}^2$, ($n=2$) because they are BPS protected.

Now, in the second step, let us further compactify the IIA action down to four dimensions on a generic six-dimensional space with volume $V_6$.
This leads to the following higher derivative action in four dimensions:
\begin{equation}
S_{4, IIA}=\int d^{4} x\sqrt{-g_{4}}\biggl( \sum_{n=1}^\infty M_P^{{1\over 2}(5-n)}e^{{1\over 6}(n-1)\phi}V_6^{{1\over 4}(n-1)} {\cal R}^n\biggr).
\end{equation}
The four-dimensional Planck mass $M_P$ is defines as usual:
\begin{equation}
M_P^2=(\alpha')^{-4}e^{-2\phi}V_6.
\end{equation}
Now it is easy to convince oneself that there exist no scaling limit of the parameters $M_P$, $e^\phi$ and $V_6$, in which only the first two terms or even only
the ${\cal R}^2$ term  survive in this expansion.
However this conclusion is only true for a generic six-dimensional compact space. In the following we like to demonstrate that for certain non-compact 
Calabi-Yau spaces  there indeed exists  a limit in which the 4-dimensional action has the form of $R+{\cal R}^2$ or even ${\cal R}^2$, with all the higher order terms being absent in a particular scaling limit.

\subsection{String theory on non-compact $CY_3$}
 
We now consider string theory on a non-compact Calabi-Yau three-fold
\cite{AMV,AFMN}.  As before, the string effective action depends on two parameters,
$e^{\phi}$ defining the loop expansion\footnote{Note that the 10-dimensional string coupling (resp. dilaton)  appears in the effective action, since we are considering
a non-compact Calabi-Yau space.}, and the string length
$l_s\sim \sqrt{\alpha'}$.  When we consider type IIA,B theories
on a background $M_{10}\,=\,M_4\,\times X_6$, 
 as shown in 
 \cite{AMV}, the effective action becomes: 
 \begin{align}
{\cal S}_{eff}&=
{1\over (2\pi)^7 l_s^8}\int_{M_4 \times   X_6}d^{10}x\sqrt{-g} e^{-2\phi} R_{(10)} + \nonumber \\
&\hspace{4mm}
{\bar \chi\over 3(4\pi)^7 l_s^2}
 \int_{M_4}\, d^{4}x\sqrt{-g_{(4)}} \left(-2\zeta(3)e^{-2\phi} 
\pm   4\zeta(2)\right)  R_{(4)} +{\cal S}_4
+\cdots \label{x}
\end{align}
 where the $\pm$-signs are related to whether we consider type IIA or B theories.
The parameter  $\bar \chi$ is given by $\bar \chi = 64\cdot 96\pi^3\, \chi$, 
 where 
 \begin{eqnarray}
 \chi =\int d^6 x\sqrt{g_6} E_6,  \label{chi}
 \end{eqnarray}
with
\begin{eqnarray}
 E_6=\frac{1}{96\pi^3}\bigg({R}_{mnrs}R^{rspq}{R_{pq}}^{mn}-2
{R^m}_{nrs}R^{rpsq}{R_{pmq}}^n \bigg).
 \end{eqnarray} 
 For a compact Ricci-flat $X_6$, $\chi$ is just the Euler number,  
 whereas for a non-compact $X_6$, this term
is not the Euler number. Boundary contributions are needed to
generate the topological invariant.  

\vskip0.3cm

In addition to the Einstein terms, there are higher order
curvature corrections to (\ref{x}) \cite{GW,Tseytlin,Kiritsis,Lu}, which are important for our discussion.  We have in particular at one-loop order:
\begin{eqnarray}
  {\cal S}_{4}= {\zeta(3)\, \omega\over  3(4\pi)^7 l_s^2}
  \int_{M_4} d^4 x\sqrt{-g_{(4)}}\, \biggl( 4R_{\mu\nu}R^{\mu\nu}-R^2\biggr), ~~~~~
  \end{eqnarray}  
where 
\begin{eqnarray}
\omega=48\int_{X_6} d^6 x \sqrt{g_6}\, R_{mnpq}R^{mnpq}, \label{omega}
\end{eqnarray}
and $(\mu, \nu,\cdots=0,\ldots,3, ~m,n,\cdots=4,\ldots,9)$.
There is also a quadratic term at tree-level but it turns out to be  just the Gauss-Bonnet combination, consistent with the fact that tree-level quadratic curvature terms should be independed of the CY moduli. 
Note that  no cosmological constant is generated because $X_6$ is a $CY$ manifold \cite{GW}. 

\vskip0.3cm

Now, taking the limits
\begin{eqnarray}
\alpha'\to 0,~~~ g_s\to 0 , \label{wcl}\, 
\end{eqnarray}
in order to suppress $\alpha'$ and string-loop corrections,
we arrive at
\begin{align}
{\cal S}_{eff}=
{2\zeta(3)\over (2\pi)^7 l_s^8}\int_{M_4 \times   X_6}\,d^{10}x 
\sqrt{-g} e^{-2\phi} R_{(10)} -
{\zeta(3)\over 3(4\pi)^7 l_s^2}
 \int_{M_4}\, d^4 x \sqrt{g_{(4)}}\,\Big{\{}e^{-2\phi} \,\bar \chi 
\ R_{(4)} 
-\omega \big{(}4R_{\mu\nu}R^{\mu\nu}-R^2 \big{)}\Big{\}} .\label{f}
\end{align}
 Therefore, string theory on a non-compact $CY_3$ induces  localized  4D terms in addition  to the bulk 10D one. This is similar to the DGP theory \cite{DGP}, where a localized 4D and a bulk 5D  Einstein term are simultaneously considered, 
 and a crossover parameter controls the regime where the effective gravity is  four- or five-dimensional. 
Here the localized 4D term dominates the bulk 10D one, as long as the 4D Planck mass $M_s/g_s$ ($M_s=1/l_s$) is much larger than the 10D Planck mass $M_s/g_s^{1/4}$ \cite{AFMN}. But this is exactly the weak string coupling limit (\ref{wcl}) $g_s\to0$. 
Let us note that the non-compact $CY_3$ is of infinite volume and therefore, the bulk dynamics is always ten-dimensional.

To explicitly demonstrate that  finite and no-vanishing 
$\chi$ and  $\omega$   
 are possible, we work out a particular case of a  non-compact $CY_3$, namely a 
complex line bundle over $\mathbb{CP}^2$. The metric for this space  can be written as \cite{Page,Lu}
\begin{eqnarray}
ds^2=\frac{1}{c^2}d\rho^2+6a^2d\sigma^2+c^2(dz+A)^2, \label{s1}
\end{eqnarray}
where 
\begin{eqnarray}
 a^2=\rho+l^2\, ~~~c^2=\frac{2\rho(\rho^2+3\rho l^2+3l^4)}{3(\rho+l^2)^2}\, 
 \end{eqnarray}
 $d\sigma^2$ is the metric of $\mathbb{CP}^2$
and $A$ is related to the K\"ahler form  on $\mathbb{CP}^2$:
\begin{eqnarray}
&&  d\sigma^2=dw^2+\frac{1}{4}\sin^2w (\sigma_1^2+\sigma_2^2)+
\frac{1}{4}\sin^2w\cos^2w \sigma_3^2,\label{cp2}\\
&&A=-\frac{3}{2}\sin^2w\, \sigma_3\,. 
  \end{eqnarray}  
  The coordinates $\rho,z$ parametrize
the $\mathbf{C}$-fiber, and to avoid conical singularities, 
$z$ takes values from $(0,2\pi)$.
The range of $\rho\in (0,\infty)$ and the range of $w\in (0,\pi/2)$.  The 1-forms $\sigma_{1,2,3}$ 
are the standard
left-invariant one forms in $SU(2)$.  This Ricci-flat metric 
contains a free parameter
$l$ that determines the minimal size of a four cycle 
diffeomorphic to $CP^2$.  In
that $l$ represents the minimal radius of this 4-cycle at $\rho=0$.
\vskip.5cm
Using this metric we can evaluate some of the quantities that 
appear in the effective actions.  In particular:
\begin{eqnarray}
&& R_{mnqp}R^{mnpq}=\frac{40(2a^2-3c^2)^2}{3a^8}=\frac{160\,l^{12}}{3(\rho+l^2)^8},\nonumber \\
&&E_6=
\frac{1}{12\pi^3}\frac{(2a^2-3c^2)^3}{a^{12}}= 
\frac{2}{3\pi^3}\frac{\,l^{18}}{(\rho+l^2)^{12}},
 \end{eqnarray}
 so that, $\omega,\chi$ defined in eqs.(\ref{omega},\ref{chi}) turn out to be
 \begin{eqnarray}
&& \omega=6(4\pi)^3 l^2,\\
&& \chi=\frac{8}{3} .
 \end{eqnarray}

Note that we have considered above only terms  up to  $R^4$. Clearly, higher order curvature terms in the bulk will be suppressed in the limit (\ref{wcl}). Namely, supergravity  corrections will be suppressed in the $\alpha'\to 0$ limit, whereas $g_s\to 0$ will suppress string loop effects. Thus, the 10D theory will be described just by the 10D Einstein term. It  is expected however, that these 10D higher  curvature terms will generate also localized 4D terms, which will contribute to localized $R^2$ or even to localized  terms of higher order like $R^4$, that  might dominate. However, a simple inspection shows that all localized $R^2$ terms arising from higher curvature terms will be suppressed in the 
$l_s\to 0$ limit. Indeed, let us consider for example a bulk $R^6$ term. The latter will give a contribution to (\ref{f}) proportional to 
\begin{eqnarray}
\omega_2\sim l_s^2
\int_{X_6} d^6 x \sqrt{g_6}\, (R_{mnpq}R^{mnpq})^2 .
\end{eqnarray}
Simply on dimensional grounds, $\omega_2\sim l_s^2/l^2$ as $l$ is the only scale in the non-compact $CY_3$. 
Therefore, such  contributions to (\ref{f}) 
vanish in the $l_s\to 0$ limit. How about localized higher order terms? For example $R^6$ will also generate a localized $R^4$ term 
proportional to $\omega$, {\it i.e.}, 
\begin{eqnarray}
 l_s^2 \int_{X_6} d^6 x \sqrt{g_6}\, R_{mnpq}R^{mnpq}
\int d^4 x\sqrt{g_4}~{\cal R}^4\sim \omega \, l_s^2 \int d^4 x\sqrt{g_4} \, {\cal R}^4,
\end{eqnarray}
so that, we will have a term 
\begin{eqnarray}
 (l_s l)^2
\int d^4 x\sqrt{g_4} ~{\cal R}^4.
\end{eqnarray}
Clearly this terms is subleading 
for finite $l$ and $l_s\to 0$ and the theory is described by (\ref{f})
in the limit (\ref{wcl}).

\subsection{The quadratic ${\cal R}^2$ action} 

It is clear then that  for weak string coupling, the bulk Planck mass is less than the 4D one and the effective theory in the limit 
(\ref{wcl}) with $\omega/l_s^2$ finite 
 is described by the action
\begin{align}
{\cal S}_{eff}=
 \int_{M_4}\, d^4 x \sqrt{g}\left(\frac{M_P^2}{2}R+
 \frac{1}{g_0^2}R_{\mu\nu}R^{\mu\nu}-\frac{1}{4g_0^2}R^2\right), \label{ffff}
\end{align}
where 
\begin{eqnarray}
M_P^2=-\frac{2\zeta(3)\bar \chi}{3(4\pi)^7l_s^2g_s^2},~~~g_0^{2}={3(4\pi)^7l_s^2\over 4\omega\zeta(3) }
\end{eqnarray}
We have ignored the Gauss-Bonnet contribution as it is a total derivative for constant $g_s$.
Note that $\bar \chi$ should be negative (as in the case for example of the 
deformed conifold). 
By using the identities (\ref{ident}) we find that 
\begin{eqnarray}
4R_{\mu\nu}R^{\mu\nu}-R^2=
{R}^{\mu\nu\rho\sigma}{R}_{\mu\nu\rho\sigma} -GB=  4 \hat R ^{\mu\nu}\hat R _{\mu\nu}=2C^{\mu\nu\rho\sigma}C_{\mu\nu\rho\sigma}-2GB+\frac{1}{3} R^2,
\end{eqnarray}
and therefore
\begin{align}
{\cal S}_{eff}&=
 \int_{M_4}\, d^4 x \sqrt{g}\left(\frac{M_p^2}{2}R+\frac{1}{g_0^2}{\hat R}^{\mu\nu}{\hat R}_{\mu\nu}\right)\nonumber\\
&=   \int_{M_4}\, d^4 x \sqrt{g}\left(\frac{M_p^2}{2}R +
\frac{1}{2g_0^2}C^{\mu\nu\rho\sigma}C_{\mu\nu\rho\sigma}+\frac{1}{12 g_0^2}R^2\right).\label{fffff}
\end{align}
We can follow now a similar procedure as described in section (\ref{r2})
to bring the theory in the Einstein frame, where, due to the conformal invariance of the Weyl tensor, the action (\ref{fffff}) turns out to be 
\begin{eqnarray}
{\cal S}_{eff}&=&
 \int_{M_4}\, d^4 x \sqrt{g}\, \left\{ \frac{M_P^2}{2} R-\frac{1}{2}(\partial \sigma)^2 -\frac{3}{4}g_0^2M_P^4\left(1-e^{-\sqrt{\frac{2}{3}}\sigma/M_P}\right)+\frac{1}{2g_0^2}C^{\mu\nu\rho\sigma}C_{\mu\nu\rho\sigma}\right\}.
 \end{eqnarray}
This is nothing else than the Starobinsky model augmented by the square of the Weyl tensor. The value of $g_0$ in this case is $g_0^2\approx 10^{-10}.$

Let us note that when the Einstein term is missing from (\ref{ffff}), we have that
\begin{eqnarray}
{\cal S}_{eff}=
 \int_{M_4}\, d^4 x \sqrt{g}\,\left( \frac{1}{2g_0^2}C^{\mu\nu\rho\sigma}C_{\mu\nu\rho\sigma}+\frac{1}{12 g_0^2}R^2\right),\label{f4}
\end{eqnarray}
which, in the Einstein  frame takes the form
   \begin{eqnarray}
{\cal S}_{eff}&=&
 \int_{M_4}\, d^4 x \sqrt{g}\, \left( \frac{1}{2g_0^2}C^{\mu\nu\rho\sigma}C_{\mu\nu\rho\sigma}+\frac{M_P^2}{2} R-\frac{1}{2}(\partial \sigma)^2 -\frac{3}{4}M_P^4\, g_0^2\right) \,. \label{f55}
 \end{eqnarray}
This  is Einstein-Weyl gravity with a massless scalar and a cosmological constant.

The spectrum of the theories (\ref{fffff}) and (\ref{f55}) contains the usual massless graviton,  a massive spin-2 ghost and a scalar (the scalaron mode). However, the ghost state should be harmless as its  appearance  can be traced here to the truncation of the string effective action in the particular framework we are discussing here.   
  
\section{Conclusions}

We have studied  here quadratic curvature gravity theories. Such theories have attracted considerable attention, among others things, due to their renormalization properties. In particular,  they offer the interesting  possibility of yielding a well
defined quantum theory, which in addition,  reduces to general relativity at long distances. We have elaborate in particular on the pure $R^2$ theory. We found that it propagates a single massless scalar and no tensor mode on a flat background. The latter appears by infrared  regulating the UV theory by considering curved backgrounds, like de Sitter or anti-de Sitter, or by introducing back the Einstein term, which dominates at long distances. Hence we observe a subtle UV/IR correspondence
in higher curvature gravity. We also discussed Newton's law in the present framework. Here we find that the extra states in the spectrum of quadratic gravity give additional contributions to the usual Newton's law. In particular, we find that there are two contributions from the spin-2 and spin-0  state  leading to an attractive force and an additional contribution from the ghost massive spin-2 state which produces a repulsive Yukawa force. Finally, we explore the possibility that quadratic gravity emerges as some particular limit of a more fundamental theory like string theory. We showed that indeed,  string theory on a non-compact $CY_3$ besides the bulk 10D gravity, it also induces a 4D localized gravity much the same way that DGP model does \cite{DGP}. In certain region of the parameter space, the localized gravity is dominating leading to a 4D effective description for the gravitational dynamics. In particular, there is a limit in which localized quadratic curvature terms are dominating leading to an effective 4D higher curvature gravity.

\vskip .5in
\noindent
{\bf \large Acknowledgment} 
\vskip .1in
We would like to 
I. Antoniadis for enlightening  discussions and M. Maggiore for correspondence on the spectrum of Einstein-Weyl gravity. 
The research of A.K. was implemented under the Aristeia II Action of the Operational Programme Education
and Lifelong Learning and is co-funded by the European Social Fund (ESF) and National Resources.
A.K. is also partially supported by European Unions Seventh Framework Programme (FP7/2007-2013)
under REA grant agreement no. 329083. 
The work of C.K.  is also supported by the CEFIPRA/IFCPAR
4104-2 project and a PICS France/Cyprus. A.R. is supported by the Swiss National Science Foundation
(SNSF), project ``The non-Gaussian Universe" (project number: 200021140236).  
D.L. likes to thank the theory group of CERN for its hospitality. This research is also supported by the Munich
Excellence Cluster for Fundamental Physics ``Origin and the Structure of the Universe"
and by the ERC Advanced Grant ``Strings and Gravity" (Grant No. 32004). The work of
C.K. is partially supported the Gay Lussac-Humboldt Research Award 2014, at the Ludwig
Maximilians University and Max-Planck-Institute for Physics.


\vskip.5in 

\end{document}